\documentclass[12pt]{article}
\usepackage{a4,amsmath,bm,amsfonts,hyperref}

\usepackage[utf8]{inputenc}
\newcommand{\de}{{\rm d}}
\newcommand{\fr}{{\textstyle\frac12}}
\newcommand{\z}{\textstyle{\frac\partial{\partial t}}}

\begin{document}
\title{Superfluid Dynamics, Equilibrium\\ Conditions,  and Centripetal Forces}
\author{Mario Liu
\\ Theoretische Physik, Universit\"{a}t T\"{u}bingen,\\ 72076 T\"{u}bingen, Germany, EC }
\date{\today}
\maketitle 
\abstract{Thermodynamics of superfluids is revisited, clarifying two points. First, the density and pressure distribution for given normal and superfluid velocities is obtained, with the finding that counter heat currents give rise to a pressure depression and a centripetal force. Second, it is shown that the ideal two-fluid hydro\-dynamics is simply an assembly of \textit{equilibrium conditions} --- expressions of entropy being maximal. 
}
\tableofcontents
\section{Introduction}
Thermodynamics of superfluids was first studied by Khalatnikov and Landau~\cite{Khal,LL6}, later by Putterman and Uhlenbeck~\cite{Putt,Putt2}. It is revisited, because 
by Galilean transforming the energy and momentum between two inertial systems, they did 
not obtain the complete kinetic contribution to the chemical potential and pressure. More specifically, they found that $\mu-\mu_s$ (where $\mu$ is the general chemical potential, and 
$\mu_s$ the one in the frame of vanishing superfluid velocity, $v^s_i=0$) is given by $\fr (v^s_i)^2-v^n_iv^s_i$. This is a pioneering result, useful for deriving the celebrated two-fluid hydrodynamics, but it lacks the information of the kinetic contribution in $\mu_s$. Only if this given, do we know the total kinetic contribution in $\mu$. The former is obtained below.  With $\mu_0$ the rest frame chemical potential, it is
\begin{align}\label{d12}
\mu_s=\mu_0(\rho,T)-\frac12\left.\frac{\partial\rho_n}{\partial\rho}\right|_Tw_i^2,\quad w_i\equiv(v^s_i-v^n_i).
\end{align}
To the best of my knowledge, this expression is new --- in spite of the vast literature in superfluid $^4$He, $^3$He, solids, and smetics, see~\cite{willi}, and references therein. (Note that $w_i$ is $-v^n_i$ in the frame of $v^s_i=0$.)

The complete motional contribution to $\mu$ enables one to calculate the density and pressure distribution in equilibrium for given velocities, especially under rotations and heat flows: First, there is a pressure decrease where heat flows, which should also be observable as a depression of the gas-liquid interface. A related effect exists under rotations, though not as pronounced: While the normal velocity $v^n_i$ gives rise to the usual centrifugal force, squeezing the density outward, any $w_i$ (whether positive or negative) diminishes this squeeze, partially drawing the density back. It acts as a centripetal force. At higher rotation rates with many superfluid vorticies present, $w_i$ averaged over a number of vorticies is small. But on smaller scales, at lower rotation rates,  $w_i$ does vary more strongly.

Maximizing the entropy, 
one obtains one Euler-Lagrange condition for each state variable, which may be referred to as \textit{equilibrium conditions}. The superfluid set of the conditions is given in Sec.~\ref{EqCond}. Two of these are 
\begin{align}
\nabla_iT=0, \quad  \z v^n_i+\nabla_i\mu=0,
\end{align}
stating that, in equilibrium, the temperature $T$ is always uniform, but the chemical potential $\mu$ is not. 
For two reasons, this seems troubling. First, $\z v^n_i\not=0$ in equilibrium; and second, the Josephson equation $\z v^s_i+\nabla_i\mu=0$ apparently relies on $\nabla_i\mu=0$ to have no superfluid acceleration in equilibrium. If $\nabla_i\mu\not=0$ sometimes, one starts to doubt the frame-independence of this important equation~\footnote{This was my initial reason for the revisit of an earlier fixation, cf.~\cite{he3-3,supersolid,he3-4,he3-5,he3-6,SC-1,SC-2}.}.  
What I eventually  found is reassuring: The two expressions $\z v^n_i+\nabla_i\mu=0$ and $\z v^s_i+\nabla_i\mu=0$ are not only compatible, they depend on each other, and may be employed with other equilibrium conditions to derive the ideal two-fluid hydrodynamics. 

\textit{Ideal dynamics} is the full one without the dissipative terms; \textit{equilibrium dynamics}, on the other hand, consists of conservation laws, the fluxes of which are assemblies of equilibrium 
conditions alone. Surprisingly, they are the same. The former does not contain any more 
information than the latter. We can use the equilibrium conditions to set up the ideal dynamics, 
then add in dissipative terms to arrive at the full dynamics. If generally true, this would systematize 
the task of setting up the macro-dynamics for any system, simplifying it considerably. Here, we 
show that  it holds true for the two-fluid dynamics. Solids, nematic liquid crystals, granular 
media and polymers will be considered in a forthcoming paper. Any full dynamics --- typically a set of nonlinear, partial differential equations --- is in its essence an expression of the relaxation toward equilibrium.

\section{Superfluid Thermodynamics}
\subsection{The  chemical potential and pressure}
In superfluids, the energy $\varepsilon$ depends on the densities of mass $\rho$, entropy $s$, momentum $g_i$, and the superfluid velocity $v^s_i=(\hbar/m)\nabla_i\varphi$. Writing 
\begin{align}\label{3}
\de\varepsilon=\mu\de\rho+T\de s+v^n_i\de g_i+j^s_i\de v^s_i,
\end{align}
defines the conjugate variables $\mu,T,v^n_i,j^s_i$. Same holds for the free energy,  $f=\varepsilon-Ts$, with
$\de f=\mu\de\rho-s\de T+v^n_i\de g_i+j^s_i\de v^s_i$.
%
The infinitesimal Galilean transformation of the velocity $\de u_i$ is given by  
\begin{align}\label{4}
{\rm d} v^s_i=\de u_i, \,\, \de g_i=\rho \de u_i,\,\, 
\de\varepsilon=\de(g_i^2/2\rho)|_\rho=g_i \de u_i,
\end{align}
holding $s,\rho=$ const. Inserting Eqs.(\ref{4}) into (\ref{3}) or $\de f$, we obtain $g_i\de u_i=\rho v^n_i\de u_i+j^s_i\de u_i$, implying
\begin{align}\label{13}
&g_i=\rho v^n_i+j^s_i,\qquad\text{and}\\\label{eq6}
&\left.\frac{\partial g_i}{\partial v^s_i}\right|_{\rho,v^n_i}=\left.\frac{\partial j^s_i}{\partial v^s_i}\right|_{\rho,v^n_i}.
\end{align}
A Maxwell relation of Eq.(\ref{3}), 
\begin{align}\nonumber
\left.\frac{\partial g_i}{\partial v^s_i}\right|_{v^n_i,\rho}=-\left.\frac{\partial j^s_i}{\partial v^n_i}\right|_{v^s_i,\rho},
\end{align}
 yields, in conjunction with Eq.(\ref{eq6}),
\begin{align}
j^s_i=\rho_s(v^s_i-v^n_i)=\rho_sw_i.\label{15}
\end{align}
The superfluid density $\rho_s$ introduced here is, to lowest order in $w_i$, independent of $w_i$. This we shall assume from here on.

Rewriting Eqs.(\ref{13},\ref{15}) as
$v^n_i=(g_i-\rho_sv^s_i)/\rho_n$, $j^s_i=(\rho v^s_i-g_i)\rho_s/\rho_n$, where $\rho_n\equiv\rho-\rho_s$, 
we infer that the free energy is
\begin{align}\label{a11}
f-f_0(\rho,T)
&=\frac{g_i^2}{2\rho_n}-\frac{\rho_s}{\rho_n}g_iv^s_i+\frac{\rho\rho_s}{2\rho_n}{(v^s_i)^2},
\end{align}
because $v^n_i={\partial f}/{\partial g_i}|_{\rho,T,v^s_i}$ and $j^s_i={\partial f}/{\partial v^s_i}|_{\rho,T,g_i}$. (This holds also for $\varepsilon-\varepsilon_0$, and the expression may be written as $\fr[\rho_s(v^s_i)^2+\rho_n (v^n_i)^2]$.)  

From Eq.(\ref{a11}), we can calculate the motional contributions to the chemical potential and pressure. 
Denoting $\mu_0\equiv{\partial f_0}/{\partial\rho}|_{T}$,  $\rho_s'\equiv{\partial \rho_s}/{\partial\rho}|_T$, with $\rho_s'+\rho_n'=1$, we find
\begin{align}\label{a12}
\mu=\mu_0(\rho,T)+\fr [\rho_s' w_i^2-(v^n_i)^2].
\end{align}
%
The pressure is $P=\rho\mu+v^n_ig_i-f=\rho\mu+Ts+v^n_ig_i-\varepsilon$. Employing Eqs.(\ref{a11},\ref{a12}) and denoting $P_0(\rho,T)\equiv\rho\mu_0-f_0$, we may write the 
pressure as\footnote{The Landau-Lifshitz expression~\cite{LL5}, $P=-\varepsilon_s+Ts+\mu_s\rho+\rho_nw_i^2$ [see Eqs.(\ref{b14}) for the definition of $\varepsilon_s$ and $\mu_s$] may be shown to be identical by simple algebra.} 
\begin{align}\label{c11}
P=P_0(\rho,T)-w_i^2 \left(\rho_s-\rho\rho_s'\right)/2.
\end{align}
Next, with $\nabla_k\varepsilon =\mu\nabla_k\rho+T\nabla_ks+v^n_i\nabla_k  g_i+j^s_i\nabla_k  v^s_i$ and $g_i= \rho v^n_i+\rho_sw_i$, see Eqs.(\ref{13},\ref{15}), the gradient of the pressure is 
\begin{align}
\label{b12}
&\nabla_k P=\rho\nabla_k\mu+s\nabla_k  T+g_i\nabla_k  v^n_i-j^s_i\nabla_k  v^s_i,
\\\label{b13}
&=\rho\nabla_k\mu+s\nabla_k  T+\fr\rho\nabla_k(v^n_i)^2-\fr\rho_s\nabla_kw_i^2.
\end{align}
Eq.(\ref{a12}) relates the distributions of density and velocities. 
For $v_{n}= v_{s}=v$, the normal fluid behavior is restored: In a rotating equilibrium, if the center is stationary, we have $\mu=\mu_0-\fr v_i^2=$ const, see the discussion below Eqs.(\ref{lokGB}) for detail. Since $v_i$ increases toward the outer rim, $\rho$ and $\mu_0(\rho)$ increase to compensate.
This is usually interpreted as a result of the universal centrifugal force pushing the density outwards: Denoting vectors with bold letters, $\bm v=\bm\Omega\times\bm r$, we have $\fr\nabla_i\bm v^2=\bm v\cdot\nabla_i\bm\Omega\times\bm r=(\bm v\times\bm\Omega)_i$.

In a superfluid, Eq.(\ref{a12}), the same holds for $\fr (v^n_i)^2$. Yet any deviation of the superfluid, $w\equiv v_{s}-v_{n}$, positive or negative, from a solid body rotation $\bm v^n=\bm\Omega\times\bm r$, even if it also rotates, operates as a centripetal force, diminishing this universality.  

Including gravitation, the chemical potential remains constant, if we add in the gravitational potential $\phi$ [see Eqs.(\ref{a12}), and Eq.(\ref{2-4c}) below]. Hence
\begin{align}\label{a22}
\nabla_iP_0/\rho&=\nabla_i\mu_0(\rho)= \nabla_i\left[\fr(v^n_i)^2 -  \fr\rho_s'w_i^2-\phi \right]
\end{align}
In a quiescent super fluid, knowing the pressure and density implies the knowledge of the function $P_0=P_0(\rho)$. Hence, measuring the density and velocity distribution under rotation is an experimental check of Eq.(\ref{a22}).  

The pressure $P$ may be measured directly in the motional fluid. To obtain the relation 
between $P$ and the two velocities, we need to replace $P_0$ in Eq.(\ref{c11}) using Eq.
(\ref{a22}). But one can more easily go to Eq.(\ref{b13}), taking $\nabla_iT=0$, $
\nabla_i\mu=-\nabla_i\phi$. Assuming incompressibility, $\rho$ and $\rho_s$ are constants and may be moved behind the gradient, implying
\begin{align}\label{c12}
P=\rho[\fr( v^n_i)^2-\phi]-\fr\rho_sw_i^2+\text{const}.
\end{align}
The first two terms show the classic behavior under rotation, with $P$ growing linearly along the axis of gravitation $-\hat z$, and quadratically with the radius. The second terms shows the superfluid modification. 

At a stationary gas-liquid interface, the form is determines by  $P=$ const. If a heat flux $w_i$ is applied, with no mass flux, the interface is depressed there, since we may then rewrite Eq.(\ref{c12}) using $\rho v^n_i+j_i^s=0$ as
\begin{align}\label{c13}
P=-\fr(\rho_s\rho_n/\rho) w_i^2-\rho\phi+\text{const}.
\end{align}

\subsection{Treatment by Landau and others}
Before presenting their results, we draw a few auxiliary conclusions. 
In the frame $v^{s}_i=0$, setting $v^{s}_i\to0$, $v^{n}_i\to (v^{n}_i-v^{s}_i)$ in Eqs.(\ref{a11},\ref{a12}), we have
\begin{align}\label{b14}
\varepsilon_s=\varepsilon_0(\rho,s)+\fr\rho_nw_i^2,
\quad 
\mu_s=\mu_0-\fr\rho_n'w_i^2,
\end{align}
implying that the general chemical potential, Eq.(\ref{a12}), may be written as 
\begin{align}\label{b15}
\mu=\mu_s-v^n_iv^s_i+\fr{(v^s_i)^2}.
\end{align}
The Galilean transformation by the velocity $u_i$, taking a system with $\hat g_i,\hat\varepsilon$ to one with $g_i,\varepsilon$,  is given by integrating Eqs.(\ref{4}) for constant $\rho$, 
\begin{align}\label{6}
\de g_i&=\rho\de u_i \,\,\succ\,\, g_i=\hat g_i+\rho u_i,\\
\de\varepsilon&=g_i\de u_i=(\hat g_i+\rho u_i)\de u_i \,\,\succ \varepsilon=\hat \varepsilon+u_i\hat g_i+\fr\rho (u_i)^2.\label{8}
\end{align}
Landau and others started from the energy in the frame $v^s_i=0$, 
\begin{align}\label{5}
\de \varepsilon_s=\mu_s\de\rho+T\de s+(v^n_i-v^s_i)\de j^n_i,\quad 
j^n_i=\rho_n(v^n_i-v^s_i),
\end{align}
and employed Eq.(\ref{8}) (with $\varepsilon_s=\hat\varepsilon$, $j^n_i=\hat g_i$) to arrive at Eq.(\ref{3}), 
obtaining the relation between $\mu$ and $\mu_s$, Eqs.(\ref{b15}).  

They also derived the equation for $v^s_i$ as
$\z v^s_i+\nabla_i[\fr (v^s_i)^2+\mu_s]=0$, though $\mu_s$ was denoted as $\mu$.
Because of Eq.(\ref{b15}), it is the same as the one obtained directly from the Josephson equation, see eg.~\cite{he3-4,PhysRevLett.38.605},
\begin{align}\nonumber
&(\z+v^n_i\nabla_i)\varphi+(m/\hbar)\mu=0, 
\\\label{1} &\text{or}\quad \z v^s_i+\nabla_i(v_k^n v_k^s+\mu)=0.
\end{align}

\subsection{Equilibrium conditions\label{EqCond}}
%
First, we derive the equilibrium conditions for a closed, quiencent  system, of given volume $\int {\rm d}^3r$, energy $\int\varepsilon_0 {\rm d^3}r$, and mass $\int \rho {\rm d^3}r$,  by maximizing  the entropy $\int s {\rm d^3}r$. This is equivalent to minimizing the energy for given entropy. 
Taking $T_L, \mu_L$ as constant Lagrange parameters, we vary $\int\varepsilon_0 {\rm d^3}r$, with $\de\varepsilon_0=\mu_0\de\rho+T_0\de s$, as
\begin{align*}\label{app2}
0=\delta\textstyle\int \left(\varepsilon_0-T_Ls-\mu_L\rho\right)\,{\rm d}^3r
=\textstyle\int\left[\left({T_0}-T_L\right)\delta
s+\left(\mu_0-\mu_L\right)\delta\rho
\right]\,{\rm d}^3r.
\end{align*}
Since $\delta s, \delta\rho$ vary independently, both brackets vanish, implying $T_0,\mu_0=$ const. Expressing these conditions locally, we have $\nabla_iT_0=0$, $\nabla_i\mu_0=0$.\footnote{Since the kinetic energy in a normal fluid, $\fr g_i^2/\rho$, depends only on $\rho$, not on $s$, there is no difference between $T_0$ and the general $T$. In superfluids, with $\rho_s$ a function of both, one needs to distinguish between $T_0$ and $T$.} 

Including gravitation, the energy is $\bar \varepsilon_0=\varepsilon_0+\rho\phi$, with the potential $\phi$ a  fixed spatial function [$\phi=\rho(z-z_0)$ on the earth's surface], implying $\bar\mu_0(\rho)\equiv\partial\bar \varepsilon_0/\partial\rho=\mu_0+\phi$. 
Varying  $\int \bar \varepsilon_0{\rm d}^3r$ under the same constraints, we obtain$\nabla_iT_0=0$ and 
\begin{equation}\label{2-4c}
\nabla_i\bar\mu_0=0,\, \text{ or }\, \nabla_i\mu_0=-\nabla_i\phi.
\end{equation}
The pressure remains unchanged, $\bar P_0=-\bar\varepsilon_0+\bar\mu_0\rho+T_0s=P_0$ for earth's potential, as does its gradient. With $\nabla_i\bar \varepsilon_0=\nabla_i\varepsilon_0 +\phi \nabla_i\rho+\rho  \nabla_i\phi$, we have 
\begin{align}\label{2-5c}
\nabla_i\bar P_0=\rho\nabla_i\bar \mu_0+s\nabla_iT_0 -\rho\nabla_i\phi=-\rho\nabla_i\phi.
\end{align}

Allowing for macroscopic motion, $v^n_i,v^s_i\not=0$, the same procedure yields the type of motion allowed in equilibrium. In essence, these are solid-body rotation, translation,  and divergence-free superflow. Minimizing the total energy $\int(\varepsilon+\rho\phi) {\rm d^3}r$, Eq.(\ref{3}), 
for given entropy $S=\int s{\rm d^3}r$,
mass $M=\int\rho{\rm d^3}r$, momentum $\bm
G=\int\bm g{\rm d^3}r$, angular momentum $\bm
L=\int(\bm r\times\bm g) {\rm d^3}r$, and booster
$\bm B=\int(\rho\bm r-\bm
gt){\rm d^3}r$ (see the explanation at the end of this section),
yields:
\begin{align}
\label{lokGB}
\nabla_i T&=0,\,\,\,\, A_{ij}\equiv\fr(\nabla_iv^n_j+\nabla_jv^n_i)=0,
\\
\nabla_i\bar\mu&+\z{v^n_i}=0,\qquad \nabla_ij_i^s=0.
\nonumber
\end{align}
The first three conditions hold also in normal fluids, except for the fact that here, $v^n_i$ takes the place of ${v_i}$, since both are given as $\partial\varepsilon/\partial g_i$. The details of the calculation is found in~\cite{3cd} which, however, does not include gravitation\footnote{ 
A cautionary remark: Including the gravitation, though the total energy $\bar w$ is conserved, the momentum, angular momentum and booster no longer are. Nevertheless,  
on the earth surface ($\phi={\cal G}z$, or $\nabla_i\phi={\cal G}\hat z_i$), $g_\perp=g_i\perp\hat z_i$,  $\ell_\parallel=\ell_i\|\hat z_i$, and the booster with $g_\perp, \ell_\|$ remain conserved; hence $\nabla_\perp\mu+\z v_\perp=0$ holds. And $\z v_z=0=\nabla_z\bar\mu$ with a  (co-moving) bottom. 
}, 
see also~\cite{Alert}. 
To obtain the fourth condition, $\nabla_ij_i^s=0$, we vary the last term in Eq.(\ref{3}), 
$\delta\int j^s_i\de \nabla_i\varphi{\rm d^3}r=\oint j^s_i\delta \varphi{\rm d^2}r-\int \nabla_i j^s_i\delta \varphi{\rm d^3}r=0$. Taking $\delta\varphi=0$ at the surface of the system (properly isolated system, no external work), and $\delta\varphi$ arbitrary within the volume, we conclude $\nabla_i j^s_i=0$. 

Clearly, the temperature $T$ is always uniform, even when the system is in motion, 
but the chemical potential $\mu$ is not; shear flow is not permitted, $A_{ij}=0$, rotation is, 
$\bm\Omega\equiv\fr\bm\nabla\times\bm v^n\not=0$.
To understand why $\z v^n_i\not=0$, consider a system rotating with $\bm \Omega$ around its center of mass, $\bm R=\bm{R_0}+\bm{\dot R} t$, that moves with a constant $\bm{\dot R}$. We have 
$\bm{v^n}=\bm{\dot R}+\bm\Omega\times(\bm r-\bm R)$, or $-\z\bm{v^n}=\bm\Omega\times\bm{\dot R}=\bm\nabla\mu\not=0$. These results are crucial for the considerations in Sec.\ref{EqDyn} to derive the equilibrium dynamics. 

In the classic book by Landau and Liftshitz on statistical mechanics,\cite{LL5} a similar consideration is given (in the sections \textit{Macroscopic Motion} and \textit{Rotating Bodies}), see also~\cite{Putt2}. Ignoring the booster, they obtained as condition $\bm\nabla\mu=0$. Yet both the angular momentum, $\boldsymbol
{\ell\equiv r\times g}$, and the booster, $\boldsymbol b\equiv\rho\boldsymbol r-\boldsymbol gt$ are locally conserved quantities -- one follows from rotational invariance, the other from Galilean invariance.  
Relativistically, both are closely related:  $\boldsymbol b$ is the zeroth component of the  4-angular momentum, the conservation of which is a result of the
Lorentz invariance: $\ell^{\alpha,\beta}=x^\alpha
g^\beta-x^\beta g^\alpha$, $x^\alpha=(ct,\boldsymbol
r)$, $g^\alpha=(\varepsilon/c\approx\rho c,
\boldsymbol g)$. Since angular momentum conservation holds
in all inertial systems, the zeroth component (that mixes with the other three under a Lorentz transformation) has to be conserved too.  Ignoring the booster is similarly wrong as ignoring one of the three components of the angular momentum.  

\subsection{Equilibrium two-fluid dynamics\label{EqDyn}}
Now we show the equivalence between equilibrium and ideal dynamics. 
The latter is given as
\begin{align}\label{26-1}
\z&  s+\nabla_i f_i=0, \quad f_i^{eq}=sv^n_i,
\\\label{26-2}
\z& \rho+\nabla_i j_i=0, 
\quad j_i=\rho v^n_i+j^s_i,
\\\label{26-3} 
\z &v^s_i+\nabla_i(v_k^n v_k^s+\hat\mu)=0, \,\,\,\hat\mu^{eq}=\mu,
\\\label{26-4} \z& g_i+\nabla_k\,\sigma_{ik}= 0,
\quad 
\sigma_{ik}^{eq}=P\delta_{ik}+g_iv_k^n+v^s_ij_k^s, 
\\\label{26-5}
\z&\varepsilon+\nabla_i Q_i=0,\,\,
Q_i^{eq}=\mu g_i+v_k^nv^s_kj^s_i+v^n_i(sT+v^n_kg_k).
\end{align}
These are five (true or quasi) conservation laws, with the fluxes specified. They have been derived~\cite{Khal,Putt2} by inserting Eqs.(\ref{26-1}--\ref{26-4}) into Eqs(\ref{3},\ref{26-5}),
\begin{align}\label{a27}
\z\varepsilon=\mu\z\rho+T\z s+v^n_i\z g_i+j^s_i\z v^s_i=-\nabla_iQ_i,
\end{align}
requiring that the four fluxes are such that they can be combined to form $\nabla_iQ_i$.
This is referred to as  the \textit{hydrodynamic procedure}. Uniqueness is certainly an issue, though more in principle than in practice. The above fluxes do satisfy the requirement, and one realizes quickly that alternative fluxes are hard to construct. The lack of rigor is compensated, and any remaining doubts eliminated, by experiments.

This paper aims to prove that given the equilibrium conditions (\ref{lokGB}), and the fact that  in equilibrium, scalar variables and parameters are stationary in the frame $\bm v^n=0$, one also arrives at the above fluxes. Employing the notation $D_tA\equiv(\z+v^n_k\nabla_k)A$  for any scalar $A$, we have, in equilibrium, 
\begin{align}\label{a31}
D_t&\rho=0,\,\, D_t s=0,\,\, D_t\rho_s=0. 
\end{align}
Starting with $D_ts=0$, we add $s\nabla_iv^n_i=0$ to yield Eq.(\ref{26-1}).  
Next, to derive Eq.(\ref{26-2}), we consider angular momentum and booster conservation,
\begin{align*}
\z\ell_m=(\boldsymbol r\times\z{\boldsymbol
g})_m=\epsilon_{mki}r_k\z
g_i=-\epsilon_{mki}r_k\nabla_j\sigma_{ij}\\ \nonumber
=-\nabla_j[\epsilon_{mki}r_k\sigma_{ij}]
+\epsilon_{mki}\sigma_{ik},
\\
\z b_i= r_i\z\rho-t\z g_i-g_i=
-r_i\nabla_jj_j+t\nabla_j\sigma_{ij}-g_i\\
 = \nabla_j(t\sigma_{ij}-r_ij_j) +j_i-g_i,
 \nonumber
\end{align*}
concluding that (both in and off equilibrium) 
\begin{equation}\label{j=g}
\sigma_{ij}=\sigma_{ji},\qquad 
j_i=g_i.
\end{equation}
Given $g_i$, Eq.(\ref{13}), we know $j_i$, Eq.(\ref{26-2}). 

For any vector $\bm B$, we denote
\begin{align}
D_t\bm B\equiv(\z+v^n_k\nabla_k-\bm\Omega\times)\bm B, \quad (\bm B\times\bm\Omega)_i=B_k\nabla_iv^n_k.
\end{align}
The second equation holds because
$(\bm B\times\bm\Omega)_i=\fr(\nabla_iv^n_k-\nabla_kv^n_i)B_k$ and because $A_{ij}=0$. 
Next, we show the validity of
\begin{align}
\label{a32}
D_t&\bm v^n=-\bm\nabla\mu,\quad D_t\bm v^s=-\bm\nabla\hat\mu.
\end{align}
We have $\z\bm v^n=D_t\bm v^n$ because $v^n_k\nabla_kv^n_i+(\bm v^n\times\bm\Omega)_i=0$, and may rewrite the equilibrium condition as $D_t\bm v^n=-\bm\nabla\mu$.
The second of Eqs.(\ref{a32}) is the same as the first of Eq.(\ref{26-3}),  because  $\nabla_i (v^n_kv^s_k)= v^n_k\nabla_iv^s_k+ v^s_k\nabla_iv^n_k =v^n_k\nabla_kv^s_i+(\bm v^s\times\bm\Omega)_i$. (Remember  $\nabla_iv^s_k= \nabla_kv^s_i\sim\nabla_i\nabla_k\varphi$.) 

To identify $\hat\mu$, we resort to the microscopic Josephson equation, 
$(\hbar/m)\z\varphi=-\mu$, which accounts for the evolution of $\varphi$ in a system at rest. If $v^n_i\not=0$, it changes to $(\hbar/m)D_t\varphi=-\mu$.  
Applying the gradient yields $\hat\mu^{eq}=\mu$.

The Josephson equation may seem an input, but is not. It can be deduced from the form of $j_i$, via energy conservation. Proceeding as prescribed by Eq.(\ref{a27}) 
we obtain (terms containing $\mu,\hat\mu^{eq}$ alone are displayed)
\begin{align}\nonumber
\nabla_i(Q_i)=\mu\nabla_i j^s_i+j^s_i\nabla_i\hat\mu^{eq}+\cdots
=\nabla_i(\mu g_i)+j^s_i\nabla_i(\hat\mu^{eq}-\mu)+\cdots
\end{align}
Only for $\hat\mu^{eq}=\mu$, is the energy $\varepsilon$ conserved, $\nabla_iQ_i=\nabla_i(\mu g_i+\cdots)$. 

Having introduced  $v^s_i$ as the new state variable, the one crucial assumption to arrive at the two-fluid dynamics is to take it transforming as a velocity. This yields $g_i$, hence $j_i$, and also the Josephson equation.

An additional important conclusion from Eq.(\ref{a32}) is that in a rotating equilibrium, if the center of mass is at rest, $\bm{\dot R}=0$, with $\bm\nabla\mu=0$, both $\bm v^s$ in the $\bm v^n$-frame and  $\bm v^n$ itself are stationary: $D_t\bm v^s=0$, $\z\bm v^n=0$. But both change with time if $\bm{\dot R}\not=0$,  even in equilibrium.

Finally, because  $\bm g=\rho \bm v^n+\bm j^s$, $D_t(\rho\bm v^n)=\bm v^nD_t\rho+\rho D_t\bm v^n$,  and $D_t\bm j^s= (\bm v^s-\bm v^n)D_t\rho_s+\rho_s D_t(\bm v^s-\bm v^n)=0$, we have
\begin{align}
\label{a34}
D_t\bm g=D_t(\rho\bm v^n)=-\rho\bm\nabla\mu.
\end{align}
Given Eq.(\ref{b12}), we note (\ref{26-4}) is the same as (\ref{a34}),
because $\nabla_iT=0$, $g_k\nabla_iv^n_k=(\bm g\times\bm\Omega)_i$,
$\nabla_k(g_iv^n_k)=v^n_k\nabla_kg_i$, $\nabla_k(v^s_ij^s_k)-j^s_k\nabla_iv^s_k=0$.
Also, the stress $g_iv^n_k+v^s_ij^s_k=\rho_nv^n_iv^n_k+\rho_sv^s_iv^s_k$ is symmetric, Eqs.(\ref{j=g}).

If we did not know in advance what the momentum flux is, these steps may appear to lack uniqueness. Yet another form of the momentum flux is quite improbable: First set $j^s_i=0$ 
to consider the stress tensor for normal fluids. 
Facing the need to rewrite $\rho\bm\nabla\mu$ of Eq.(\ref{a34})  as $\nabla_
j\sigma_{ij}$, there is little choice other than switching to the pressure, $\nabla_iP$. Rewriting the remaining term $v^n_k\nabla_kg_i$ in $D_tg_i$ as $\nabla_k(g_iv^n_k)$ is a familiar step, with $g_iv^n_k=\rho v^n_iv_k^n$ symmetric. Going back to superfluid, only $g_iv^n_k+v^s_ij^s_k$ is symmetric, and $\nabla_k(v^s_ij^s_k)=j^s_k\nabla_iv^s_k$ cancels  the same term in the pressure gradient. 

Energy conservation $\z\varepsilon+\nabla_i Q_i=0$ is not independent. To calculate $Q_i$, one proceeds as in Eq.(\ref{a27}), now keeping every term. Since all fluxed are known, it is easily done. That all terms combine to form the divergence of $Q_i$, is clearly an added argument against alternative fluxes. 

To add gravitation, we write  $\z g_i+\nabla_k\,\sigma_{ik}=-\rho\nabla_i\phi$, Eqs.(\ref{2-4c},\ref{2-5c}), implying  $\z\varepsilon+\nabla_iQ_i=-\rho v^n_i\nabla_i\phi$, since  $\z\varepsilon=v^n_i\z g_i+\cdots=-\rho v^n_i\varepsilon\nabla_i\phi+\cdots$.

\section{Dissipative fluxes}
Off equilibrium, the entropy production does not vanish, $\z  s+\nabla_i f_i=R/T\not=0$, 
neither do the equilibrium conditions, 
\[\nabla_i T,\, A_{ij},\,  \nabla_ij_i^s,\,   \z{v^n_i}+\nabla_i\mu\not=0,\] 
and are now referred to as \textit{thermodynamic forces}.  
As they completely characterize a macro state, both in and off equilibrium, we may expand $R$ in them, $R=\kappa(\nabla_iT)^2+\cdots$,  to second order. There are no constant or linear terms, as $R$ vanishes and is minimal in equilibrium. 
At the same time, dissipative fluxes appear, $f=f^{eq}+f^D$, $\hat\mu=\hat\mu^{eq}+\hat\mu^D$, $\sigma_{ij}=\sigma_{ij}^{eq}+\sigma_{ij}^D$, where $j_i^D=j_i-j_i^{eq}$ vanishes identically. 
Proceeding again as in Eq.(\ref{a27}), including now the dissipative fluxes, we find  an expanded $Q_i$, and
\begin{align}
R=j^D_i\nabla_i\mu+f^D_i\nabla_iT+\hat\mu^D\nabla_ij^s_i+\sigma_{ij}^DA_{ij},
\end{align}
enabling us to conclude that $f^D$, $\hat\mu^D$, $\sigma_{ij}^D$ are linear combinations of $\nabla_iT$, $\nabla_ij^s_i$, $A_{ij}$ as prescribed by the Onsager rules --- with $f^D_i\sim\nabla_iT$, we indeed have  $R=\kappa(\nabla_iT)^2+\cdots$. 
Since $j^D_i\equiv0$, neither  $\z{v^n_i}+\nabla_i\mu$ nor $\nabla_i\mu$ serves as a fourth thermodynamic force.

The basic insight of the above consideration is the central role of equilibrium conditions, both in and off equilibrium. The two-fluid hydrodynamics is simply the addition of equilibrium dynamics plus thermodynamic forces pushing the system back to equilibrium. There are indications that this is true of many more systems, including solid, nematic/smectic liquid crystals, polymeric solutions and granular media.


\begin{thebibliography}{99}

\bibitem{Khal}
Isaac~M. Khalatnikov.
\newblock {\em An introduction to the theory of superfluidity}.
\newblock CRC Press, 2018.

\bibitem{LL6}
Lev~Davidovich Landau and Evgeny~Mikhailovich Lifshitz.
\newblock Fluid mechanics, volume 6 of course of theoretical physics.
\newblock {\em Course of theoretical physics/by LD Landau and EM Lifshitz}, 6,
  1987.

\bibitem{Putt}
Seth Putterman and G.~E. Uhlenbeck.
\newblock Thermodynamic equilibrium of rotating superfluids.
\newblock {\em The Physics of Fluids}, 12(11):2229, 1969.

\bibitem{Putt2}
S.~J. Putterman.
\newblock {\em Superfluid Hydrodynamics}.
\newblock North Holland, 1974.

\bibitem{willi}
Johannes Hofmann and Wilhelm Zwerger.
\newblock Hydrodynamics of a superfluid smectic.
\newblock {\em J. Stat. Mech. Theory Exp.}, 2021(3):033104.

\bibitem{3cd}
Peter Kost{\"a}dt and Mario Liu.
\newblock Three ignored densities, frame-independent thermodynamics, and broken
  {G}alilean symmetry.
\newblock {\em Physical Review E}, 58(5):5535, 1998.

\bibitem{he3-3}
Mario Liu.
\newblock Hydrodynamics of he 3 near the a transition.
\newblock {\em Physical Review Letters}, 35(23):1577, 1975.

\bibitem{supersolid}
Mario Liu.
\newblock Two possible types of superfluidity in crystals.
\newblock {\em Physical Review B}, 18(3):1165, 1978.

\bibitem{he3-4}
Mario Liu and M.~C. Cross.
\newblock Broken spin-orbit symmetry in superfluid he 3 and b-phase dynamics.
\newblock {\em Physical Review Letters}, 41(4):250, 1978.

\bibitem{he3-5}
Mario Liu and M.~C. Cross.
\newblock Gauge wheel of superfluid he 3.
\newblock {\em Physical Review Letters}, 43(4):296, 1979.

\bibitem{he3-6}
Mario Liu.
\newblock Broken relative symmetry and the hydrodynamics of superfluid 3he.
\newblock {\em Physica B+ C}, 109:1615--1628, 1982.

\bibitem{SC-1}
Mario Liu.
\newblock Rotating superconductors and the frame-independent london equation.
\newblock {\em Physical review letters}, 81(15):3223, 1998.

\bibitem{SC-2}
Yimin Jiang and Mario Liu.
\newblock Rotating superconductors and the london moment: thermodynamics versus
  microscopics.
\newblock {\em Physical Review B}, 63(18):184506, 2001.


\bibitem{LL5}
Lev~Davidovich Landau, Evgenii~M. Lifshitz, and L.~P. Pitaevskii.
\newblock Statistical physics, part i, 1980.

\bibitem{PhysRevLett.38.605}
Chia-Ren Hu and Wayne~M. Saslow.
\newblock Hydrodynamics of $^{3}\mathrm{He}$-$a$ with arbitrary textures.
\newblock {\em Phys. Rev. Lett.}, 38:605--609, Mar 1977.

\bibitem{Alert}
Mario Liu.
\newblock Thermodynamics and constitutive modeling.
\newblock In Itai Einav and Eleni Gerolymatou, editors, {\em ALERT Doctoral
  School 2018; Energetical Methods in Geomechanics}, pages 3--42. The Alliance
  of Laboratories in Europe for Education, Research and Technology, 2018.

\end{thebibliography}
\end{document}